\documentstyle[psfig,11pt]{article}
\newcommand{\BS}{{{\it Beppo}SAX}\ \ignorespaces}
\newcommand{\ltsima} {$\; \buildrel < \over \sim \;$}
\newcommand{\gtsima} {$\; \buildrel > \over \sim \;$}
\newcommand{\lta} {\lower.5ex\hbox{\ltsima}}
\newcommand{\gta} {\lower.5ex\hbox{\gtsima}}
\begin{document}
\vspace{1.0cm}
{\Large \bf \BS OBSERVATIONS OF THE BLACK HOLE CANDIDATES LMC X--1 AND LMC X--3}

\vspace{1.0cm}

A. Treves$^1$, M. Galli$^1$, F. Haardt$^1$, T. Belloni$^2$, L. Chiappetti$^3$, 
D. Dal Fiume$^4$, F. Frontera$^{4,5}$, E. Kulkeers$^6$, L. Stella$^7$

\vspace{1.0cm}
$^1${\it Dipartimento di Scienze, Universit\`a dell'Insubria/Polo di Como, 
Italy}\\
$^2${\it Astronomical Institute, Amsterdam, the Netherland}\\
$^3${\it IFCTR, CNR, Milano, Italy}\\
$^4${\it ITESRE, CNR, Bologna, Italy}\\
$^5${\it Dipartimento di Fisica, Universit\`a di Ferrara, Italy}\\
$^6${\it Nuclear and Astrophysics Laboratory, University of Oxford, 
United Kingdom}\\
$^7${\it Osservatorio di Roma, Monteporzio, Italy}\\

\vspace{0.5cm}

\section*{ABSTRACT}
We describe \BS observations of the black hole candidates LMC X--1 and 
LMC X--3 performed in Oct. 1997. Both sources can be modelled by a multicolor 
accretion disk spectrum, with temperature $\sim 1$ keV. However, there 
is some evidence that a thin emitting component coexists with the thick 
disk at these temperatures. In the direction of LMC X--1, we detected 
a significant emission above 10 keV, which we suspect originates from the 
nearby source PSR 0540-69. For LMC X--1, 
we estimate an absorbing column density of $\simeq 6\times 10^{21}$ cm$^{-2}$, 
which is almost ten times larger than that found for LMC X--3. In both sources, 
we find no indication of emission or absorption features whatsoever.

\section{INTRODUCTION}
LMC X--3 and LMC X--1 are two luminous persistent X--ray binaries. LMC X--3 has
an orbital period of 1.7 days and is one of the most secure black hole
candidates ($M_{\rm X} \simeq 10M_{\odot}$). For LMC X--1 there is still some 
uncertainty regarding
the optical counterpart. The most probable optical candidate has a period of
4.23 d, which implies a companion (hole) mass of $\simeq 5 M_{\odot}$ 
(Cowley et al. 1995).
  
The two objects have soft X--ray spectra reminiscent of Cyg X--1 in the
high/soft state. A high energy tail extending above 10 keV is detected in both
systems (Ebisawa, Mitsuda \& Inoue 1989; Treves et al. 1990). 

We observed the two sources with \BS in Oct. 1997, which allowed us to
construct simultaneous spectra in a broad energy band (0.1--100 keV). All the
four narrow field instruments (LECS 0.1--4 keV, MECS 1.8--10 keV,
HPGSPC 7--70 keV, PDS 12--150 keV) 
performed nominally; the MECS functioning with two
units. Here we are presenting some preliminary results on the X--ray spectral 
distribution. HPGSPC data analysis is not discussed in this paper.

Our observations correspond to a low intensity state of LMC X--3, while LMC
X--1 appears stable on month/year time scale, as apparent from the
the XTE--ASM light curves. 

\begin{table*}
\begin{tabular}{cccccccc}
\multicolumn{8}{l}{{TABLE 1: Observation Log}}\\
&&&&&&&\\ 
\hline
\hline
&&&&&&&\\
\multicolumn{1}{c}{Source Name}
&\multicolumn{3}{c}{Exposure Time (ks)}&
&\multicolumn{3}{c}{Count rate (cts/s)}\\
& 
\multicolumn{1}{c} {LECS} & 
\multicolumn{1}{c} {MECS} &
\multicolumn{1}{c} {PDS} & &
\multicolumn{1}{c} {LECS} &
\multicolumn{1}{c} {MECS} &
\multicolumn{1}{c} {PDS}\\
&&&&&&&\\
\hline
&&&&&&&\\
{LMC X--1} & {14.4} & {38.2} & {43.3}& & 
{3.26$\pm{0.02}$} & {2.67$\pm{0.01}$} & {0.21$\pm{0.05}$}\\
&&&&&&&\\
{LMC X--3} & {17.4} & {39.5} & {38.0}& &
{3.37$\pm{0.01}$} & {2.19$\pm{0.01}$} & {0.03$\pm{0.05}$}\\
&&&&&&&\\
\hline
&&&&&&&\\
&&&&&&&\\
\multicolumn{8}{l}{Note: MECS counts are for one unit. PDS counts 
are for four units.}
\end{tabular}
\end{table*}

\begin{table*}
\begin{tabular}{cccccccc}
\multicolumn{8}{l}{{TABLE 2: Spectral Fits: Disk Spectrum$^{a}$ 
+ Free--Free}}\\
&&&&&&&\\ 
\hline
\hline
&&&&&&&\\
\multicolumn{1}{c}{Source Name}
&\multicolumn{1}{c}{$N_{\rm H}$}
&\multicolumn{1}{c}{$kT_{\rm in}$}
&\multicolumn{1}{c}{N$_{\rm diskbb}^{b}$}
&\multicolumn{1}{c}{$kT_{\rm brems}$}
&\multicolumn{1}{c}{N$_{\rm bremss}^{c}$}
&\multicolumn{1}{c}{F$^{d}_{[2-10]{\rm keV}}$}
&\multicolumn{1}{c}{$\chi^2/dof$}\\
&{($10^{22}$ cm$^{-2}$)} & {(keV)} & &{(keV)} & & 
& \\
&&&&&&&\\
\hline
&&&&&&&\\
{LMC X-1} & 
{$0.61^{+0.03}_{-0.03}$} & 
{$0.85^{+0.01}_{-0.02}$} & 
{$45.1^{+7.0}_{-8.5}$} & 
{$2.39^{+0.25}_{-0.19}$} &
{$0.15^{+0.04}_{-0.04}$}&  
{$3.25$} &
{$190/214$}\\
&&&&&&&\\  
{LMC X-3} & 
{$0.056^{+0.005}_{-0.004}$} & 
{$1.05^{+0.01}_{-0.02}$} & 
{$19.4^{+1.7}_{-1.2}$} &  
{$2.01^{+0.42}_{-0.90}$} &
{$0.036^{+0.008}_{-0.009}$}&    
{$2.71$} &
{$217/214$}\\
&&&&&&&\\
\hline
&&&&&&&\\
&&&&&&&\\
\multicolumn{8}{l}{Note: errors are 90\% confidence level for one parameter.}\\
&&&&&&&\\
\multicolumn{8}{l}{ $^a$~
${dN \over dE}= {8\pi \over 3}{R_{\rm in}\over D^2} \cos{\theta} \,
\int_{T_{\rm out}}^{T_{\rm in}} (T/T_{\rm in})^{-11/3} B(E,T) dT/T_{\rm in}$}\\
&&&&&&&\\
\multicolumn{8}{l}{ $^b$~ $[(R_{\rm in}/{\rm km})/
(D/10{\rm kpc})]^2 \cos{\theta}$ } \\
&&&&&&&\\
\multicolumn{8}{l}{ $^c$~   
${3.02\times 10^{-15} \over 4\pi D^2} \int n_e n_I\, dV$}\\
&&&&&&&\\
\multicolumn{8}{l}{ $^d$~Flux unit: (10$^{-10}$ erg/cm$^2$/s)}
\end{tabular}
\end{table*}

\section{\BS SPECTRA}

Data reduction followed the standard procedure.
Total exposure times and count rates are reported in Table 1.
Results of the spectral analysis described below are reported in Table 2.

\begin{itemize}
\item[LMC X--1:]

In the field of view of the MECS and PDS there is the source PSR 0540-69. 
In the $[2-10]$ keV band, the pulsar is fainter than LMC X--1 by a factor 
$\sim$ 10. The pulsar spectrum is rather hard, and a power--law fit of the 
MECS data yields an energy index $\Gamma=2.01\pm{0.07}$. In principle, 
the pulsar emission could account of (part of) the signal detected in the PDS. 
To test this hypothesis, we first fitted separately the 
LECS+MECS data of LMC X--1 with an absorbed disk blackbody, obtaining a 
totally unacceptable fit ($\chi^2/dof > 3$). The inclusion 
of an optically thin free--free component greatly improves the fit 
($\chi^2/dof=0.89$; see Table 2 and Fig. 1). 
We did not find any narrow emission 
or absorption feature. The column density is quite large, 
$N_{\rm H} \simeq 6.5 \times 10^{21}$ cm$^{-2}$. 
Then, we fitted the PDS data alone. They are well 
represented by a hard power--law, which is completely consistent, in 
terms of flux and 
spectral index, with the extrapolation in the PDS energy range of the MECS 
spectrum of the pulsar. We
therefore suspect that the pulsar is responsible of the high energy emission 
detected in the PDS.


\begin{figure}
\centerline{\psfig{figure=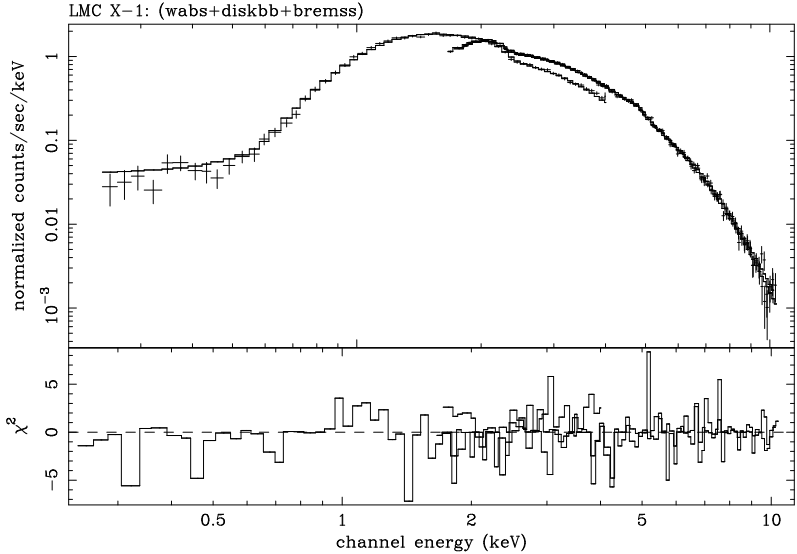,height=9.0cm,angle=-90}}
\caption{ LMC X--1: 
count spectrum and contribution to $\chi^2$ when data are fitted with 
a {\tt wabs[diskbb+bremss]} model.}
\end{figure}

\item[LMC X--3:]
In this source, we did not find any positive detection in the PDS. 
A variable hard X--ray tail was observed in the PDS in only one of the two 
one--month spaced \BS science verification phase observations performed 
in 1996, while in the other it had an upper limit comparable 
to ours (Siddiqui et al. 1998). 
As in the case of LMC X--1, a fit with an absorbed multicolor disk spectrum 
showed evidence of features in the 0.8--2 keV range ($\chi^2$ null probability 
$\lta 0.1$\%). The inclusion of a 
free--free component reduced the $\chi^2$ to an 
acceptable value (see Table 2 and Fig. 2). 
Again, we did not find any narrow 
emission or absorption feature above 2 keV. 
The absorbing column lies in the range reported by 
previous X--ray observations ($N_{\rm H}\simeq 5.7 
\times 10^{20}$ cm$^{-2}$, Treves et al. 1988).
\end{itemize}

\begin{figure}
\centerline{\psfig{figure=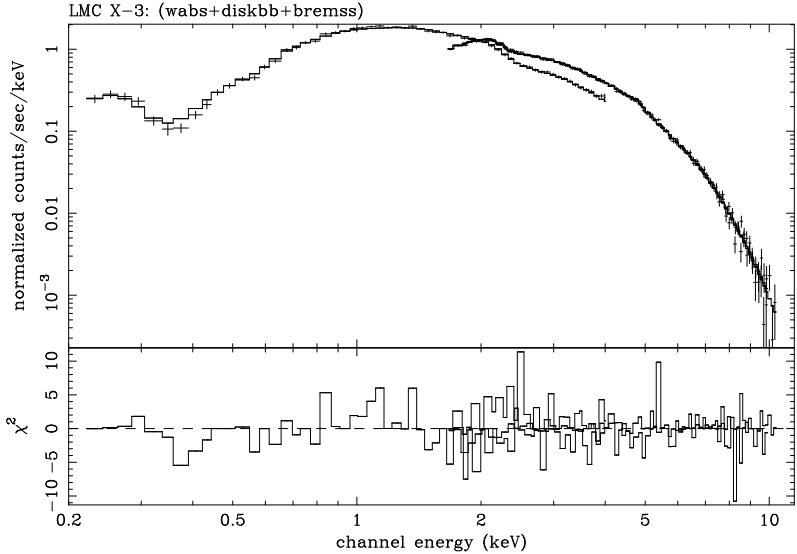,height=9.0cm,angle=-90}}
\caption{LMC X--3: 
count spectrum and contribution to $\chi^2$ when data are fitted with 
a {\tt wabs[diskbb+bremss]} model.}
\end{figure}

\section{CONCLUSIONS}

Our analysis shows that in LMC X--1 and LMC X--3 an optically thick 
accretion disk coexists with an optically thin, X--ray emitting gas, with 
comparable temperatures. This may indicate that the thermal emission from 
the innermost region of the accretion disk is modified by electron scattering. 
In LMC X--1 direction there is also indication of a component at much 
higher energy, responsible 
for the power--law emission in the PDS, which however we suspect arises from a
nearby pulsar. Previous claims of the presence of a hard tail in LMC X--1 
were based on non imaging instruments, where confusion with the nearby 
pulsar can not be excluded. LMC X--1 is 
found to be severely absorbed at low energies. 
Theoretical interpretations of the results are in progress and will be 
presented in a forthcoming paper.

\section{REFERENCES}
\vspace{-5mm}
\begin{itemize}
\setlength{\itemindent}{-8mm}
\setlength{\itemsep}{-1mm}

\item[] 
Cowley, A.P., Schmidtke, P.C., Anderson, A.L., and McGrath, T.K., 1995, 
PASP, 107, 145.

\item[] 
Ebisawa, K., Mitsuda, K., and Inoue, H., 1989, PASJ, 41, 519.

\item[]
Siddiqui, H., et al., 1998, in prep.

\item[] 
Treves, A., et al., 1988, ApJ, 325, 119.

\item[] 
Treves, A., et al., 1990, ApJ, 364, 266.

\end{itemize}

\end{document}